\documentclass[preprint, 1p]{elsarticle}

\usepackage{amsmath,amssymb,bm}
\usepackage{lineno}
\usepackage{color}

\usepackage{multirow}
\usepackage{multicol}

\journal{Physics Letters B}

\begin{document}

\begin{frontmatter}

\title{Intrinsic spin distributions in multinucleon transfer reactions}

\author[ITP]{D. D. Zhang}

\author[PKU,Zagreb]{D. Vretenar\corref{mycorrespondingauthor}}
\ead{vretenar@phy.hr}

\author[Zagreb]{T. Nik\v si\' c}

\author[PKU]{P. W. Zhao\corref{mycorrespondingauthor}}
\ead{pwzhao@pku.edu.cn}

\author[PKU]{J. Meng\corref{mycorrespondingauthor}}
\ead{mengj@pku.edu.cn}

\address[ITP]{Institute of Theoretical Physics, Chinese Academy of Science, Beijing 100871, China}
\address[PKU]{State Key Laboratory of Nuclear Physics and Technology, School of Physics, Peking University, Beijing 100871, China}
\address[Zagreb]{Physics Department, Faculty of Science, University of Zagreb, 10000 Zagreb, Croatia}

\cortext[mycorrespondingauthor]{Corresponding authors}

\begin{abstract}
Time-dependent covariant density functional theory (TD-CDFT) combined with angular momentum projection is developed and applied to study multinucleon transfer (MNT) reactions, with a focus on the intrinsic angular momentum distributions of the final fragments. Using the illustrative reaction $^{40}$Ca + $^{208}$Pb across a range of impact parameters, we find that the MNT process generates broad distributions of intrinsic spins. These distributions arise from the conversion of relative orbital angular momentum into intrinsic spin due to frictional interactions between the colliding nuclei. Additionally, mutual information (entanglement Shannon entropy) is employed to analyze correlations between the intrinsic spins of the fragments.
\end{abstract}


\end{frontmatter}

\section{Introduction}

The synthesis of superheavy nuclei (SHN) offers valuable insights into nuclear structure and reaction dynamics. Of particular interest are neutron-rich SHN near the island of stability, which are predicted to exhibit exotic structural and chemical properties. These unique characteristics have attracted widespread attention not only in nuclear physics but also in chemistry and related fields.

Currently, fusion reactions represent the sole viable method for synthesizing SHN. However, all SHN produced via fusion to date are neutron-deficient. Furthermore, the synthesis of new SHN through fusion is hindered by extremely low cross-sections, necessitating either radioactive beams of higher intensity or more optimized projectile-target combinations to advance progress in this field.

Multinucleon transfer (MNT) reactions represent a promising pathway for synthesizing neutron-rich heavy and superheavy nuclei, making them a top priority at many nuclear reaction facilities \cite{GSI_RotaruINCREASE2022, IMP_HuangNPR2017, France_GalesPPNP2007, Italy_MijatovicFOP2022, CorradiNIMPRB2013}. Theoretical studies suggest that MNT reactions could be particularly effective for producing neutron-rich nuclei near the $N = 126$ shell closure \cite{ZagrebaevPRL2008}. Experimental evidence supports this prediction: the MNT reaction $^{136}$Xe + $^{198}$Pt yields significantly larger cross-sections than fragmentation reactions with a $^{208}$Pb target, especially on the neutron-rich side \cite{WatanabePRL2015}. This finding highlights the potential of MNT reactions for accessing neutron-rich nuclei in the $N = 126$ region.

Additionally, theoretical work has suggested that MNT reactions may offer advantages for superheavy nucleus synthesis due to enhanced shell stabilization effects \cite{ZagrebaevPRC2006, ZagrebaevJPG2007, ZagrebaevPRC2008, ZagrebaevPRC2011}. A recent experimental breakthrough—the production of the neutron-rich nucleus $^{241}$U via the MNT reaction $^{238}$U + $^{198}$Pt—further demonstrates the viability of this approach \cite{NiwasePRL2023}. However, despite these advances, no superheavy nuclei have yet been synthesized through MNT reactions, primarily due to the complex mass and angular distributions of the reaction products~\cite{HuangNPR2017}.

While mass distributions in nuclear reactions have been extensively investigated~\cite{SimenelPRL2010, SekizawaPRC2013, ZhaokaiPRC2016, WenPRC2019, JiangPRC2020, WuPLB2022, ChenpenghuiPRC2022, FengzhaoqingPRC2023, LiaoPRC2024, ZhaotianliangPRC2024, ZhangPRC2024, ZDDPRC2024, ChengLiPRC2024, ZhangfengshouPRC2025, FengPingPRC2025, OcalPRC2025, GaoPRC2025}, their associated angular momentum distributions remain relatively unexplored~\cite{LiaoPRR2023, ScampsPRC2024, LiaoPRC2025, LePRC2025}. This knowledge gap stems from the complex dynamics involved, 
where part of the total angular momentum is converted into intrinsic spins of final fragments ($ \vec S_{L}$, $\vec S_{H}$) and the remaining relative orbital angular momentum ($\vec \Lambda$):

\begin{equation}
\vec J = \vec S_{L} + \vec S_{H} + \vec \Lambda.
\end{equation}

The mechanism of angular momentum redistribution in this process remains poorly understood~\cite{ScampsPRC2024}. Furthermore, due to the contribution of $\vec \Lambda$, the spin of one fragment does not uniquely determine the other fragment's spin. A key unresolved question is whether the intrinsic spins of the two fragments exhibit any correlation.

The angular momentum distribution and correlations between fragments have been extensively studied in both spontaneous and induced fission of even-even nuclei~\cite{WilsonNature2021, RandrupPRL2021, VogtPRC2021, BulgacPRL2021, BulgacPRL2022, ScampsPRC2023}. However, the mechanism of angular momentum generation in fission fragments and their potential correlations have remained unresolved for over four decades, with competing models offering different explanations. A recent breakthrough in experimental measurements of fission fragments' intrinsic spins revealed no significant correlation between the spins of complementary fragments, suggesting that angular momentum generation occurs predominantly after scission~\cite{WilsonNature2021}. This finding has stimulated vigorous discussions and renewed interest in fission dynamics~\cite{RandrupPRL2021, VogtPRC2021, BulgacPRL2021, BulgacPRL2022, ScampsPRC2023}.

The situation in MNT reactions presents additional complexity. Unlike fission, MNT reactions typically involve non-zero total angular momentum, originating from both the intrinsic spins of the projectile and target nuclei and the orbital angular momentum inherent in non-central collisions. This multifaceted angular momentum composition makes the analysis of fragment spins and their correlations in MNT reactions particularly challenging.

In a recent study of Ref.~\cite{ScampsPRC2024}, a microscopic study of spin transfer in near-barrier nuclear reactions, based on nuclear TD-DFT, has shown that the transfer of nucleons and neck formation can significantly affect the transfer of spin through tangential friction.
In this Letter, the intrinsic spin distributions of fragments and their correlations in the MNT reaction $^{40}$Ca + $^{208}$Pb are investigated using microscopic time-dependent covariant density functional theory (TD-CDFT)~\cite{RenPRC2020, RenPLB2020, RenPRC2022, RenPRC2022-chiral, RenPRL2022, LiPRC2023, LiPRC2024-fission, LiPLB2024}. Since the final reaction products typically lack well-defined quantum numbers (e.g., intrinsic spins), we develop an extended TD-CDFT framework incorporating angular momentum projection to resolve fragment spins. We analyze the mechanism of spin generation in MNT reactions and quantify spin-spin correlations between fragments using mutual information (entanglement Shannon entropy).

\section{Theoretical framework}
The time evolution of the reaction system is modelled by the TD-CDFT. The evolution of single-nucleon wave functions $\psi_k(\bm{r})$ is governed by the Dirac equation
\begin{equation}\label{eq: Dirac equation}
	i\hbar\frac{\partial}{\partial t}\psi_k(\bm{r},t)=[\bm{\alpha}\cdot(\hat{\bm{p}}-\bm{V})+V^0+\beta(m+S)]\psi_k(\bm{r},t).
\end{equation}
Here, $\bm{\alpha}$ and $\beta$ are the Dirac matrices, $\hat{\bm{p}}$ is the momentum operator, and $m$ is the mass of the nucleon. The scalar potential $S$ and vector potential $V^{\mu}$ are determined self-consistently by the time-dependent densities and currents. Further details can be found in Ref.~\cite {RenPRC2020}. 

After the collision, the conservation of the total angular momentum requires
\begin{equation}
\vec J=\vec S_{\text{PLF}}+\vec S_{\text{TLF}}+\vec \Lambda,
\end{equation}
where $\vec J$ is the total angular momentum of the reaction system, $\vec S_{\text{PLF}}$ and $\vec S_{\text{TLF}}$ are the intrinsic spins of the projectile-like fragment (PLF) and target-like fragment (TLF) after they separate, respectively. $\vec \Lambda$ is the relative orbital angular momentum between the two fragments. Because the mean-field wave functions of PLF and TLF are not eigenstates of the intrinsic spin operator, rotational symmetry is restored by angular momentum projection. 

The angular momentum projection operator reads:
\begin{equation}
\hat{P}^S_{MK}=\frac{2S+1}{16\pi^2}\int d\Omega(D^S_{MK}(\Omega))^*\hat{R}(\Omega),
\end{equation}
where $\Omega=(\alpha,\beta,\gamma)$ represents the three Euler angles. $D^S_{MK}(\Omega)$ is the Wigner function, where $M$ and $K$ are projections of the intrinsic spin $\vec S$ on the symmetry axis in the laboratory and body-fixed frames, respectively. $\hat{R}(\Omega)$ is the rotation operator:
\begin{equation}
\hat{R}(\Omega)=e^{-i\alpha\hat{S}_z}e^{-i\beta\hat{S}_y}e^{-i\gamma\hat{S}_z},
\end{equation}
with ${\hat{S}}_{y,z}$ representing the intrinsic spin operators along $y$- and $z$-axis. The probability distribution for the PLF or TLF with an intrinsic spin $S$ and projection $K$ is computed from the expression:
\begin{align}
P(S,K)=&\langle\Psi|(\hat{P}^S_{MK})^{\dagger}\hat{P}^S_{MK}|\Psi\rangle=\langle\Psi|\hat{P}^S_{KK}|\Psi\rangle\nonumber\\
=&\frac{2S+1}{16\pi^2}\int_0^{2\pi}d\alpha\int_0^{\pi}\sin\beta d\beta\int_0^{4\pi}d\gamma(D^S_{KK})^*\nonumber\\
&\langle\Psi|e^{-i\alpha\hat{S}_z}e^{-i\beta\hat{S}_y}e^{-i\gamma\hat{S}_z}|\Psi\rangle.
\end{align}
$|\Psi\rangle$ is the wave function of PLF or TLF, which can be obtained by multiplying the single-nucleon wave functions with the Heaviside step function in the three-dimensional (3D) lattice space,
\begin{align}
\Psi(\bm{r})=\frac{1}{\sqrt{A!}} \det\{\Theta_V(\bm{r})\psi_k(\bm{r})\},
\end{align}
where $A$ is the total number of nucleons. The Heaviside function $\Theta_V(\bm{r})$ divides the space into the volume $V$ in which the PLF (TLF) is located at a given time, and the complementary space 
\begin{align}
\Theta_V(\bm{r})=\begin{cases}
1\quad \text{if}\ \bm{r}\in V,\\
0\quad \text{if}\ \bm{r}\notin V.
\end{cases}
\end{align}
Since the intrinsic spin operator is a one-body operator, the overlap $\langle\mathrm{\Psi}|e^{-i\alpha\hat{S}_z}e^{-i\beta\hat{S}_y}e^{-i\gamma\hat{S}_z}|\mathrm{\Psi}\rangle$ can be expressed through rotation of single-particle states $|\psi_k\rangle$ 
\begin{equation}
\langle\mathrm{\Psi}|e^{-i\alpha\hat{S}_z}e^{-i\beta\hat{S}_y}e^{-i\gamma\hat{S}_z}|\mathrm{\Psi}\rangle=\det\{\langle\psi_k|\psi_k^{rot}\rangle\},
\end{equation}
with $|\psi_k^{rot}\rangle=\hat{R}(\alpha,\beta,\gamma)|\psi_k\rangle$.

After obtaining the spin distributions, correlations between different spin observables can be quantified by evaluating their entanglement Shannon entropy (mutual information). The entanglement Shannon entropy between two observables $A$ and $B$ is defined by the following expression~\cite{MaPPNP2018}: 
\begin{equation}
I(A,B)=H(A)+H(B)-H(A,B),
\end{equation}
where $H(A), H(B)$, and $H(A,B)$ are Shannon entropies computed from the corresponding distributions,
\begin{subequations}
\begin{equation}
H(A)=-\sum_i P(A_i)\ln P(A_i),
\end{equation}
\begin{equation}
H(B)=-\sum_i P(B_i)\ln P(B_i),
\end{equation}
\begin{equation}
H(A,B)=-\sum_{ij} P(A_i,B_j)\ln P(A_i,B_j).
\end{equation}
\end{subequations}
If $A$ and $B$ are independent, then $P(A, B)=P(A)P(B)$ and the mutual information $I(A, B)$ vanishes. Conversely, two observables are correlated if the entanglement Shannon entropy differs from zero. 

As a representative case study, we examine the MNT reaction $^{40}{\rm Ca}+{}^{208}{\rm Pb}$ at laboratory energy $E_{\rm lab} = 249$ MeV. Our calculations employ the relativistic density functional PC-PK1~\cite{ZhaoPRC2010}, both in static (CDFT)~\cite{RenPRC2017, LiPRC2020, ZDDPRC2022, ZDDIJMPE2023, XuPRC2024, XuPRL2024, XuPLB2024} and time-dependent covariant density functional theory (TD-CDFT) calculations~\cite{RenPRC2020, RenPLB2020, RenPRC2022, RenPRC2022-chiral, RenPRL2022, LiPRC2023, LiPRC2024-fission, LiPLB2024}.

For the stationary solution, we use a 3D lattice with dimensions $N_x\times N_y\times N_z = 26\times 26\times 26$ and a mesh spacing of 0.8 fm, while dynamical calculations employ an asymmetric lattice of $60\times 26\times 60$ with the same spacing. The Dirac equation (Eq. \eqref{eq: Dirac equation}) is solved numerically using a predictor-corrector method with a time step of 0.2 fm/$c$. The initial configuration positions the projectile and target nuclei 24 fm apart on the lattice, assuming Rutherford trajectories prior to collision. The time evolution continues until either two well-separated fragments emerge with a center-of-mass distance exceeding 20 fm or 24 fm (impact-parameter dependent), or the system evolves into a composite nucleus (fusion). 
In angular momentum projection, for the integration over angles, we employ trapezoidal integration with uniform discretized points in the $\alpha$ and $\gamma$ directions, while using non-uniform Gauss-Legendre integration in the $\beta$ direction.
The Euler angles are discretized with 
$30 \times 30 \times 30$ mesh points for the PLF and 
$40 \times 40 \times 40$ for the TLF, ensuring sufficient resolution for spin distribution analysis. To obtain the projection along the $K$-axis, which is defined as the line connecting the centers of mass of the PLF and TLF, we rotate the nuclei to let the $K$-axis be parallel to the $z$-axis before projecting on angular momentum.

\section{Results and discussion}
In a recent study \cite{LiPRC2024}, we investigated the MNT reaction between $^{40}{\rm{Ca}}$ with $^{208}{\rm{Pb}}$ at  $E_{\rm{lab}} = 249$ MeV, analyzing the von Neumann entropies, total entanglement between fragments, nucleon-number fluctuations, and Shannon entropy for the nucleon-number observable. Using identical model parameters to those employed here, it was shown that fusion occurs when the impact parameter $b$ is smaller than 4.64 fm, while nucleon transfer becomes negligible for $b\geq  10$ fm. Consequently, the present study adopts this same impact parameter range.

Because both the projectile $^{40}$Ca and target $^{208}$Pb have zero initial intrinsic spin, the total angular momentum of the system equals the initial relative orbital angular momentum. For collisions occurring in the $x$-$z$ plane, this results in a non-zero orbital angular momentum component along the $y$-axis: $\Lambda_y=b\sqrt{2\mu E_{\text{cm}}}$, determined by the non-vanishing impact parameter $b$. Throughout the dynamical evolution, TD-CDFT calculations explicitly conserve the expectation value of the total angular momentum.

\begin{figure}[htbp]
  \centering
  \includegraphics[width=0.75\textwidth]{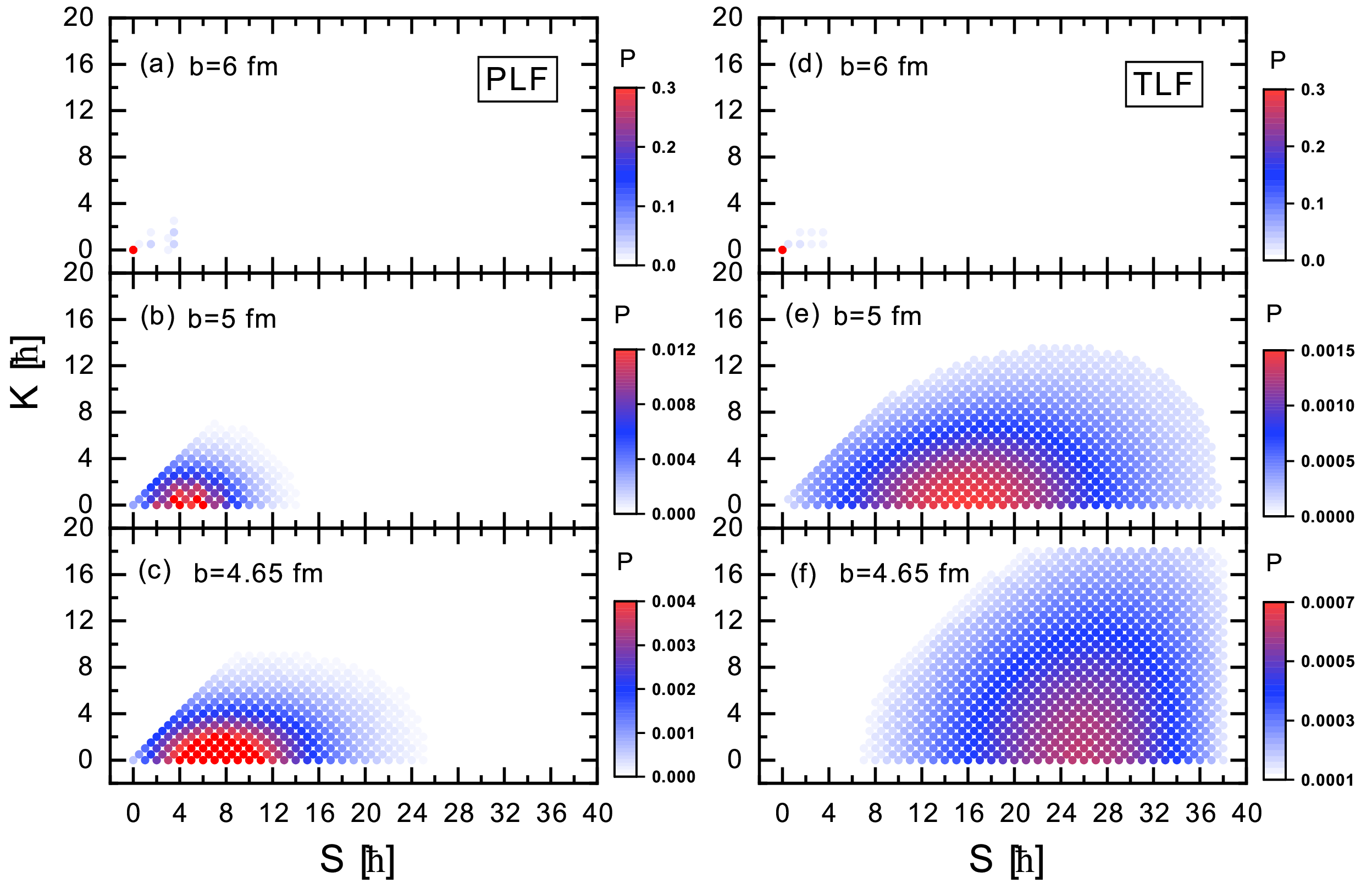}
  \caption{\label{fig1} (Color online). The intrinsic spin distributions with $K\geq0$, for the projectile-like fragment (PLF) and target-like fragment (TLF), at impact parameters $b=6$ fm, $5$ fm, and $4.65$ fm.}
\end{figure}

Taking the final fragments of the reactions at impact parameters $b=6$ fm ($\Lambda=109.9~\hbar$), $5$ fm ($\Lambda=91.6~\hbar$), and $4.65$ fm ($\Lambda=85.2~\hbar$) as examples, the intrinsic spin distributions of the PLF and the TLF are shown in Fig.~\ref{fig1}. 
Due to the $x$-signature symmetry, the following relation holds: $P(S, -K)=P(S, K)$ and we display only the distributions for $K\ge 0$.
At $b \geq 6$ fm, $^{40}$Ca and $^{208}$Pb barely come into contact, and the average number of transferred nucleons is close to zero \cite{LiPRC2024}. As shown in Fig.~\ref{fig1}(a) and (d), the spins of both fragments are nearly zero. 

As the impact parameter decreases to $5$ fm, nucleon transfer occurs between the PLF and TLF, resulting in average particle numbers of $Z\approx18, N\approx23$ for the PLF and $Z\approx84, N\approx123$ for the TLF. Angular momenta in both fragments are generated and exhibit a broad distribution, as shown in Fig.~\ref{fig1}(b) and (e), indicating the conversion of relative orbital angular momentum into intrinsic spin. This behavior arises from frictional forces—sliding and rolling friction—between the fragments, where sliding corresponds to rigid rotation of the di-nuclear system, while rolling involves one fragment rotating over the other~\cite{ScampsPRC2024, Bass1980}. 

As discussed in Ref.~\cite{ScampsPRC2024}, the ratio of the average spins, $S_\text{PLF}/S_\text{TLF}$, is expected to be  $\approx A_{\text{PLF}}^{1/3}/A_{\text{TLF}}^{1/3}$ for sticking equilibrium, and $\approx A_{\text{PLF}}^{5/3}/A_{\text{TLF}}^{5/3}$ for rolling equilibrium. Our calculations yield a ratio of $6.73/19.08=0.35$ at $b=5$ fm, which lies between the sticking (0.58) and rolling (0.07) limits. This suggests that both friction mechanisms contribute to the dynamics. Notably, the TLF exhibits a larger intrinsic spin than the PLF, contrasting with fission systems where the lighter fragment typically carries higher spin.

As the impact parameter decreases further to $4.65$ fm, the average nucleon composition shifts to $Z\approx23, N\approx32$ for the PLF and $Z\approx79, N\approx114$ for the TLF. The fragment spins increase significantly, with broader distributions emerging in both the $S$ and $K$ directions. This demonstrates a more substantial conversion of angular momentum into the fragments' intrinsic degrees of freedom.

\begin{figure}[htbp]
  \centering
  \includegraphics[width=0.75\textwidth]{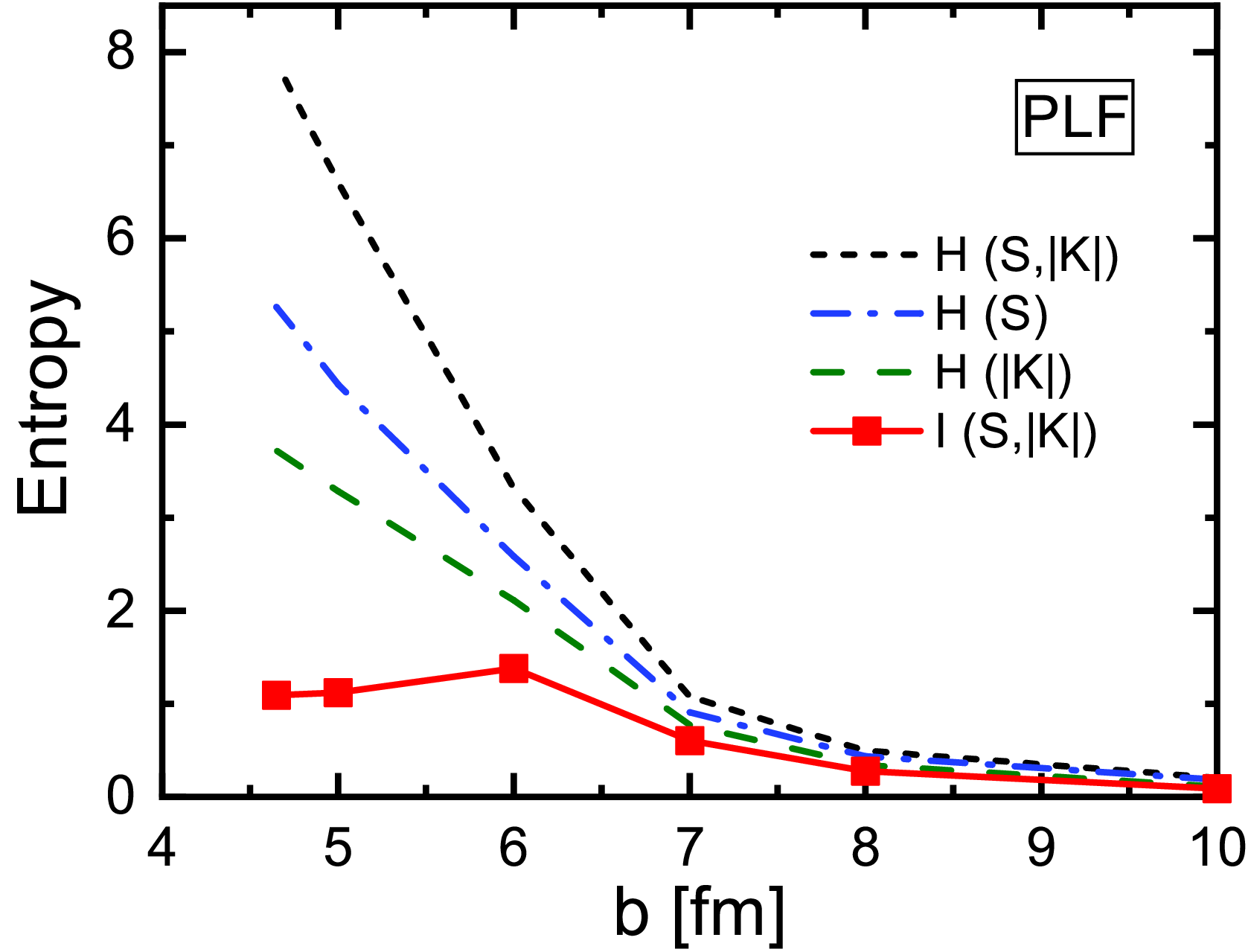}
  \caption{\label{fig2} (Color online). The Shannon entropies $H(S,|K|)$, $H(S)$, and $H(|K|)$, as functions of the impact parameter $b$, computed from the distributions $P(S, |K|)$, $P(S)$, and $P(|K|)$ for the PLF. The values of the entanglement Shannon entropy $I(S,|K|)$ are denoted by the square symbol. The lines are used to guide the eye.  }
\end{figure}

To assess the validity of entanglement Shannon entropy as a measure of correlations between two observables, we first examine the relationship between $S$ and $|K|$, noting the symmetry $P(S, K)=P(S, -K)$. Figure~\ref{fig2} displays the Shannon entropies derived from the distributions $P(S, |K|)$, $P(|K|)$, and $P(S)$ in the PLF, along with the mutual information. The marginal distributions of the $S$ and $|K|$ are obtained via $P(S)=\sum_K P(S, K)$ and $P(|K|)=\sum_SP(S, |K|)$. At large impact parameters (e.g., $b=10$ fm), the entanglement Shannon entropy is nearly zero, as evidenced by $P(S=0,|K|=0)=P(S=0)\times P(|K|=0)=1$, indicating negligible correlation. However, as the impact parameter decreases, the entanglement Shannon entropy rises to approximately 
$1.5$, reflecting enhanced correlations between $S$ and $|K|$. This trend is further supported by the fragment spin distributions at $b=5$ fm and $b=4.65$ fm, shown in Fig.~\ref{fig1}(b) and (c). 
If $S$ and $|K|$ were independent, the joint distribution would factorize as $P(S, |K|)=P(S)P(|K|)$, subject to the geometric constraint $S\geq |K|$. However, the distributions in Figs.~\ref{fig1}(b) and (c) deviate from this expectation, indicating a correlation between $S$ and $|K|$. In Fig.~\ref{fig2}, the mutual information initially follows the trend of the Shannon entropies, increasing for smaller impact parameters $b$.  However, for $b < 6$ fm, $I(S,|K|)$ no longer mirrors the Shannon entropies and even exhibits a slight decline. A similar trend is observed for the corresponding quantities in the TLF. This suggests that as nucleon transfer between the nuclei intensifies, the correlation between $S$ and $|K|$ weakens. As shown in Fig.~\ref{fig1}(b) and (c), while more orbital angular momentum is converted into fragment spins, their projections along the symmetry axis ($|K|$) remain confined to relatively small values. This implies that the spins are predominantly oriented perpendicular to the collision plane. Consequently, the mutual information $I(S, |K|)$ does not increase with the Shannon entropies at smaller impact parameters.

To evaluate the correlations between the intrinsic spins of two fragments, the joint spin distribution $P(S_{\text{PLF}}, S_{\text{TLF}})$  is required for mutual information calculations, where $S_{\text{PLF}}$ and $S_{\text{TLF}}$ are the spins of the PLF and TLF, respectively. However, obtaining this distribution would involve a computationally expensive projection over six Euler angles, making it infeasible for practical reaction analysis. As a more tractable alternative, we first consider the projection of spin along a single axis. The resulting distribution of fragment spins along a chosen direction can then be expressed as:
\begin{align}
P(K_{\text{PLF}},K_{\text{TLF}})=&\langle\Psi| \hat{P}^{(\text{P})}_{K_{\text{PLF}}}\hat{P}^{(\text{T})}_{K_{\text{TLF}}}|\Psi\rangle,
\end{align}
where the symbol $(\text{P})$ or $(\text{T})$ denotes the projection operator that only acts on the wave function of the PLF or TLF, respectively, defined as:
\begin{equation} 
\hat{P}_{k}=\frac{1}{4\pi}\int_0^{4\pi} d\theta e^{i\theta k}e^{-i\theta\hat{S}_m},
\end{equation}
where $\hat{S}_m$ is the intrinsic spin operator along a certain $m$-axis, $k=K_{\text{PLF}}$ ($K_{\text{TLF}}$) is the spin projection along the $m$-axis for the PLF (TLF).

\begin{figure}[htbp]
	\centering
	\includegraphics[width=0.75\textwidth]{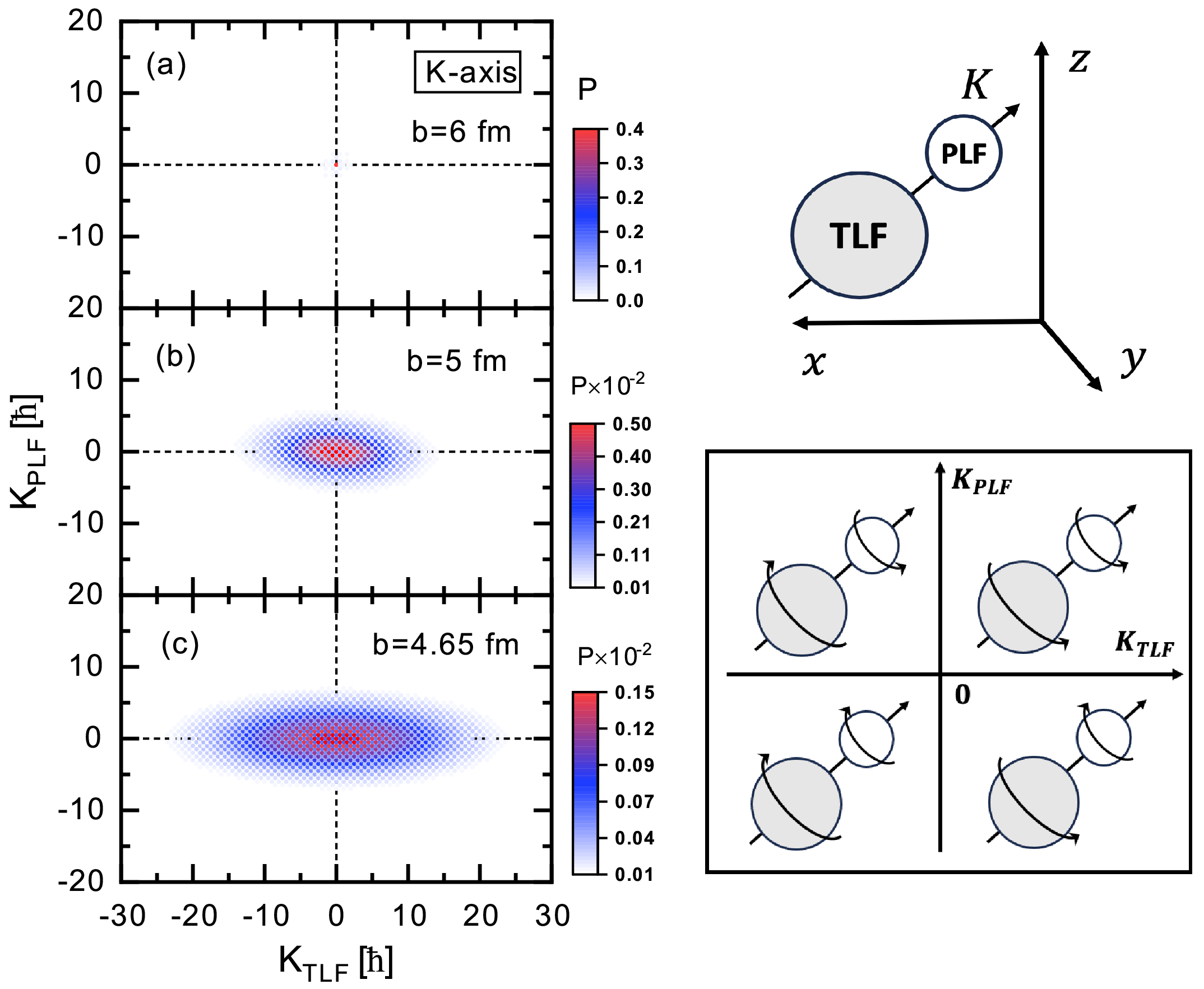}
	\caption{\label{fig3} (Color online). The distributions of the intrinsic spin projections $K_{\text{PLF}}$ and $K_{\text{TLF}}$ along the $K$-axis, defined as the line connecting the centers of mass of the PLF and TLF. }
\end{figure}

\begin{figure}[htbp]
	\centering
	\includegraphics[width=0.75\textwidth]{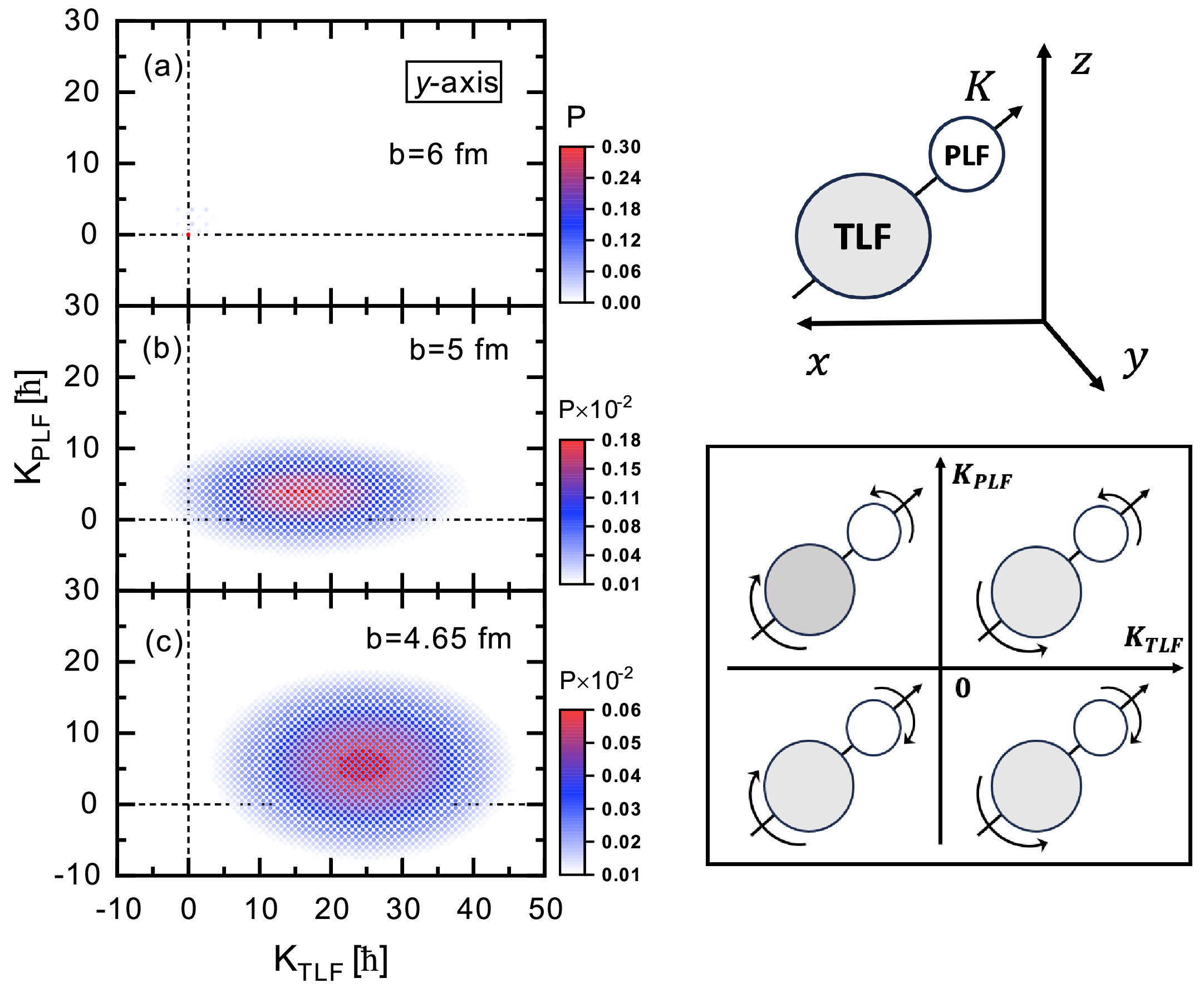}
	\caption{\label{fig4} (Color online). Same as Fig.~\ref{fig3} but for the $y$-axis perpendicular to the collision plane.}
\end{figure}

We analyze two principal directions, as illustrated schematically in the upper right panel of Fig.~\ref{fig3}: (1) the $K$-axis defined as the line connecting the centers of mass of the two fragments, and (2) the $y$-axis, perpendicular to the reaction plane. 
The distributions of the intrinsic spin projections along the $K$- and $y$-axes are presented in Figs.~\ref{fig3} and \ref{fig4}, respectively. 


In the $K$-direction, the average spins of both fragments vanish. The distributions $P(K_{\text{PLF}}, K_{\text{TLF}})$ peak at the origin for impact parameters $b = 6$ fm, $5$ fm, and $4.65$ fm, as shown in Figs.~\ref{fig3}(a)-(c). This implies zero average spins for both PLF and TLF along $K$, which consequently requires the average relative orbital angular momentum to also vanish, since the total angular momentum in the $K$-direction is conserved and remains zero throughout the reaction. These findings are consistent with the observation that the fragment velocities lie entirely within the $x$-$z$ plane.
Notably, the TLF exhibits a broader distribution of spin projections along $K$ compared to the PLF. Additionally, distinct collective modes emerge: as illustrated in the lower right panel of Fig.~\ref{fig3}, fragments rotate in the same direction when $K_{\text{PLF}}K_{\text{TLF}} > 0$, while exhibiting counterrotation when $K_{\text{PLF}}K_{\text{TLF}} < 0$.

In the $y$-direction, the total angular momentum has a non-zero average value. Figures~\ref{fig4}(b) and (c) reveal that intrinsic spins are predominantly distributed in the region where both $K_{\text{PLF}}>0$ and $K_{\text{TLF}}>0$, with this tendency becoming particularly pronounced at smaller impact parameters (e.g., $b=4.65$ fm). 
These observations align with earlier findings demonstrating the conversion of relative angular momentum into intrinsic spin.
Of course, the spins could also be mainly distributed in the region with both $K_{\text{PLF}}<0$ and $K_{\text{TLF}}<0$, depending on whether the initial orbital angular momentum is oriented along the positive or negative direction of the $y$-axis.
The lower right panel of Fig.~\ref{fig4} illustrates the corresponding collective modes: fragments co-rotate about the $y$-axis when $K_{\text{PLF}}K_{\text{TLF}}>0$, while counter-rotating when $K_{\text{PLF}}K_{\text{TLF}}<0$. 
These modes are similar to known fission modes, including tilting, twisting, bending, and wriggling modes~\cite{MorettoPRC1980, MorettoNPA1989}.

\begin{figure}[htbp]
  \centering
  \includegraphics[width=0.75\textwidth]{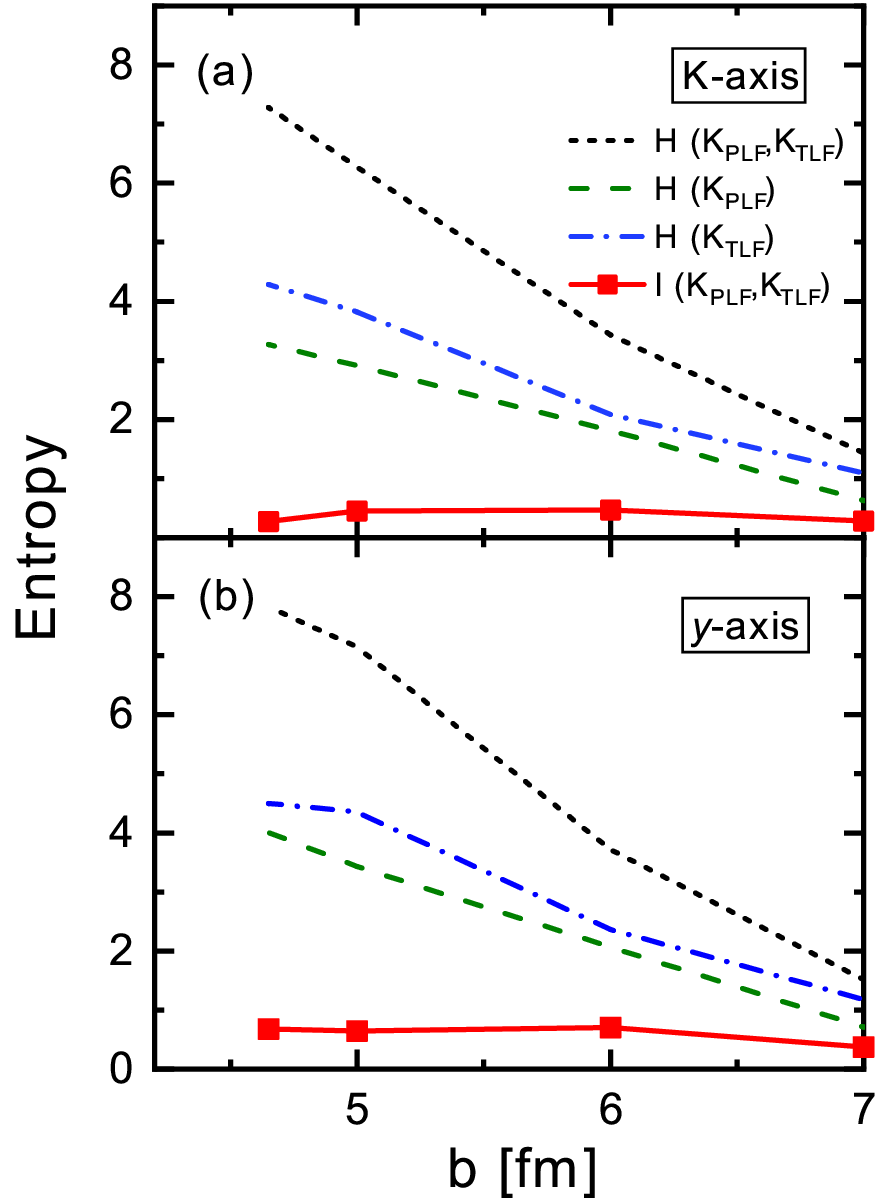}
  \caption{\label{fig5} (Color online). (a) The Shannon entropies $H(K_{\text{PLF}}, K_{\text{TLF}})$, $H(K_{\text{PLF}})$, and $H(K_{\text{TLF}})$, as functions of the impact parameter $b$, computed from the distributions $P(K_{\text{PLF}}, K_{\text{TLF}})$, $P(K_{\text{PLF}})$, and $P(K_{\text{TLF}})$ for spin projections along the $K$-axis, defined as the line connecting the centers of mass of the PLF and TLF. The values of the entanglement Shannon entropy $I(K_{\text{PLF}}, K_{\text{TLF}})$ are denoted by the square symbol. The lines are used to guide the eye. The corresponding results for projections on the $y$-axis are shown in the panel $(b)$.}
\end{figure}

Finally, we quantify the spin correlations between the two fragments along the $K$- and $y$-axes using entanglement Shannon entropy. Figures~\ref{fig5}(a) and (b) present the Shannon entropies $H(K_{\text{PLF}}, K_{\text{TLF}})$, $H(K_{\text{PLF}})$, and $H(K_{\text{TLF}})$ as functions of impact parameter $b$, derived from the respective spin projection distributions $P(K_{\text{PLF}}, K_{\text{TLF}})$, $P(K_{\text{PLF}})$, and $P(K_{\text{TLF}})$. Although non-zero, the entanglement Shannon entropies for both axes remain remarkably small and show negligible dependence on $b$, demonstrating that spin-spin correlations between the fragments are exceptionally weak.

\section{Summary}

We have developed and applied a time-dependent covariant density functional theory incorporating angular momentum projection to study fragment intrinsic spin distributions and their correlations in multinucleon transfer (MNT) reactions. For the specific case of $^{40}$Ca + $^{208}$Pb at $E_{\text{lab}}=249$ MeV, we have computed fragment spin distributions across impact parameters from 4.65 fm to 10 fm - a range where the experimental MNT cross-sections are well reproduced. Our calculations reveal that:
\begin{itemize}
\item Broad intrinsic spin distributions emerge during the MNT process, generated through the conversion of relative orbital angular momentum to intrinsic spins via frictional coupling between fragments.
\item Entanglement Shannon entropy analysis of the PLF shows correlations between $S$ and $|K|$ that become significant at smaller impact parameters but, as nucleon transfer between the nuclei intensifies, the mutual information between $S$ and $|K|$ weakens.
\item Projections along the $K$-axis (inter-fragment axis) and $y$-axis (perpendicular to the reaction plane) reveal various collective rotational modes.
\item Mutual information quantification of inter-fragment spin correlations shows non-zero but weak correlations along both $K$ and $y$ directions.

\end{itemize}

\section*{Acknowledgments}
The authors are grateful to G. Scamps and S. G. Zhou for useful discussions. 
This work is supported by the National Key Research and Development Program of China (Contract No. 2023YFA1606503, No. 2024YFE0109801, No. 2024YFE0109803 and No. 2024YFA1612600), the National Natural Science Foundation of China (Grants No. 12447101, No. 12375118, No. 12435008, No. W2412043, No. 12435006, No. 12475117, No. 12141501, and No. 11935003), and the CAS Strategic Priority Research Program (Grant No. XDB34010103, and No. XDB0920102). 
This work has been supported in part by the High-End Foreign Experts Plan of China, the National Key Laboratory of Neutron Science and Technology (Grant No. NST202401016), by the project ``Implementation of cutting-edge research and its application as part of the Scientific Center of Excellence for Quantum and Complex Systems, and Representations of Lie Algebras'', PK.1.1.02, European Union, European Regional Development Fund, and by the Croatian Science Foundation under the project Relativistic Nuclear Many-Body Theory in the Multimessenger Observation Era (IP-2022-10-7773).
This work is supported by the Postdoctoral Fellowship Program and China Postdoctoral Science Foundation under Grant Number BX20250170.
The results described in this paper are obtained on the High-performance Computing Cluster of ITP-CAS and the Sc-Grid of the Supercomputing Center, Computer Network Information Center of the Chinese Academy of Sciences.


\begin{thebibliography}{61}
\expandafter\ifx\csname natexlab\endcsname\relax\def\natexlab#1{#1}\fi
\providecommand{\url}[1]{\texttt{#1}}
\providecommand{\href}[2]{#2}
\providecommand{\path}[1]{#1}
\providecommand{\DOIprefix}{doi:}
\providecommand{\ArXivprefix}{arXiv:}
\providecommand{\URLprefix}{URL: }
\providecommand{\Pubmedprefix}{pmid:}
\providecommand{\doi}[1]{\href{http://dx.doi.org/#1}{\path{#1}}}
\providecommand{\Pubmed}[1]{\href{pmid:#1}{\path{#1}}}
\providecommand{\bibinfo}[2]{#2}
\ifx\xfnm\relax \def\xfnm[#1]{\unskip,\space#1}\fi
\bibitem[{Rotaru et~al.(2022)Rotaru, Amanbayev, Balabanski, Benyamin,
  Constantin, Dickel, Gr{\"o}f, Mardor, Miskun, Nichita, Pla{\ss},
  Scheidenberger, Sp{\u a}taru, and State}]{GSI_RotaruINCREASE2022}
\bibinfo{author}{A.~Rotaru}, \bibinfo{author}{D.~Amanbayev},
  \bibinfo{author}{D.~L. Balabanski}, \bibinfo{author}{D.~Benyamin},
  \bibinfo{author}{P.~Constantin}, \bibinfo{author}{T.~Dickel},
  \bibinfo{author}{L.~Gr{\"o}f}, \bibinfo{author}{I.~Mardor},
  \bibinfo{author}{I.~Miskun}, \bibinfo{author}{D.~Nichita},
  \bibinfo{author}{W.~R. Pla{\ss}}, \bibinfo{author}{C.~Scheidenberger},
  \bibinfo{author}{A.~Sp{\u a}taru}, \bibinfo{author}{A.~State},
  \bibinfo{journal}{Nucl. Instrum. Methods Phys. Res. B} \bibinfo{volume}{512}
  (\bibinfo{year}{2022}) \bibinfo{pages}{83--90}.
  \DOIprefix\doi{10.1016/j.nimb.2021.11.018}.
\bibitem[{Huang et~al.(2017)Huang, Tian, Wang, Wang, Gan, Liu, Yang, Ma, and
  Zhou}]{IMP_HuangNPR2017}
\bibinfo{author}{W.~X. Huang}, \bibinfo{author}{Y.~L. Tian},
  \bibinfo{author}{Y.~S. Wang}, \bibinfo{author}{J.~Y. Wang},
  \bibinfo{author}{Z.~G. Gan}, \bibinfo{author}{Z.~Liu},
  \bibinfo{author}{J.~Yang}, \bibinfo{author}{X.~W. Ma}, \bibinfo{author}{X.~H.
  Zhou}, \bibinfo{journal}{Nucl. Phys. Rev.} \bibinfo{volume}{34}
  (\bibinfo{year}{2017}) \bibinfo{pages}{409--413}.
  \DOIprefix\doi{10.11804/NuclPhysRev.34.03.409}.
\bibitem[{Gales(2007)}]{France_GalesPPNP2007}
\bibinfo{author}{S.~Gales}, \bibinfo{journal}{Prog. Part. Nucl. Phys.}
  \bibinfo{volume}{59} (\bibinfo{year}{2007}) \bibinfo{pages}{22--31}.
  \DOIprefix\doi{10.1016/j.ppnp.2006.12.021}.
\bibitem[{Mijatovi{\'c}(2022)}]{Italy_MijatovicFOP2022}
\bibinfo{author}{T.~Mijatovi{\'c}}, \bibinfo{journal}{Front. Phys.}
  \bibinfo{volume}{10} (\bibinfo{year}{2022}).
  \DOIprefix\doi{10.3389/fphy.2022.965198}.
\bibitem[{Corradi et~al.(2013)}]{CorradiNIMPRB2013}
\bibinfo{author}{L.~Corradi}, et~al., \bibinfo{journal}{Nucl. Instrum. Methods
  Phys. Res. B} \bibinfo{volume}{317} (\bibinfo{year}{2013})
  \bibinfo{pages}{743}. \DOIprefix\doi{10.1016/j.nimb.2013.04.093}.
\bibitem[{Zagrebaev and Greiner(2008)}]{ZagrebaevPRL2008}
\bibinfo{author}{V.~Zagrebaev}, \bibinfo{author}{W.~Greiner},
  \bibinfo{journal}{Phys. Rev. Lett.} \bibinfo{volume}{101}
  (\bibinfo{year}{2008}) \bibinfo{pages}{122701}.
  \DOIprefix\doi{10.1103/PhysRevLett.101.122701}.
\bibitem[{Watanabe et~al.(2015)Watanabe, Kim, Jeong, Hirayama, Imai, Ishiyama,
  Jung, Miyatake, Choi, Song, Clement, de~France, Navin, Rejmund, Schmitt,
  Pollarolo, Corradi, Fioretto, Montanari, Niikura, Suzuki, Nishibata, and
  Takatsu}]{WatanabePRL2015}
\bibinfo{author}{Y.~X. Watanabe}, \bibinfo{author}{Y.~H. Kim},
  \bibinfo{author}{S.~C. Jeong}, \bibinfo{author}{Y.~Hirayama},
  \bibinfo{author}{N.~Imai}, \bibinfo{author}{H.~Ishiyama},
  \bibinfo{author}{H.~S. Jung}, \bibinfo{author}{H.~Miyatake},
  \bibinfo{author}{S.~Choi}, \bibinfo{author}{J.~S. Song},
  \bibinfo{author}{E.~Clement}, \bibinfo{author}{G.~de~France},
  \bibinfo{author}{A.~Navin}, \bibinfo{author}{M.~Rejmund},
  \bibinfo{author}{C.~Schmitt}, \bibinfo{author}{G.~Pollarolo},
  \bibinfo{author}{L.~Corradi}, \bibinfo{author}{E.~Fioretto},
  \bibinfo{author}{D.~Montanari}, \bibinfo{author}{M.~Niikura},
  \bibinfo{author}{D.~Suzuki}, \bibinfo{author}{H.~Nishibata},
  \bibinfo{author}{J.~Takatsu}, \bibinfo{journal}{Phys. Rev. Lett.}
  \bibinfo{volume}{115} (\bibinfo{year}{2015}) \bibinfo{pages}{172503}.
  \DOIprefix\doi{10.1103/PhysRevLett.115.172503}.
\bibitem[{Zagrebaev et~al.(2006)Zagrebaev, Oganessian, Itkis, and
  Greiner}]{ZagrebaevPRC2006}
\bibinfo{author}{V.~I. Zagrebaev}, \bibinfo{author}{Y.~T. Oganessian},
  \bibinfo{author}{M.~G. Itkis}, \bibinfo{author}{W.~Greiner},
  \bibinfo{journal}{Phys. Rev. C} \bibinfo{volume}{73} (\bibinfo{year}{2006})
  \bibinfo{pages}{031602}. \DOIprefix\doi{10.1103/PhysRevC.73.031602}.
\bibitem[{Zagrebaev and Greiner(2007)}]{ZagrebaevJPG2007}
\bibinfo{author}{V.~Zagrebaev}, \bibinfo{author}{W.~Greiner},
  \bibinfo{journal}{J. Phys. G: Nucl. Part. Phys.} \bibinfo{volume}{34}
  (\bibinfo{year}{2007}) \bibinfo{pages}{2265}.
  \DOIprefix\doi{10.1088/0954-3899/34/11/004}.
\bibitem[{Zagrebaev and Greiner(2008)}]{ZagrebaevPRC2008}
\bibinfo{author}{V.~Zagrebaev}, \bibinfo{author}{W.~Greiner},
  \bibinfo{journal}{Phys. Rev. C} \bibinfo{volume}{78} (\bibinfo{year}{2008})
  \bibinfo{pages}{034610}. \DOIprefix\doi{10.1103/PhysRevC.78.034610}.
\bibitem[{Zagrebaev and Greiner(2011)}]{ZagrebaevPRC2011}
\bibinfo{author}{V.~I. Zagrebaev}, \bibinfo{author}{W.~Greiner},
  \bibinfo{journal}{Phys. Rev. C} \bibinfo{volume}{83} (\bibinfo{year}{2011})
  \bibinfo{pages}{044618}. \DOIprefix\doi{10.1103/PhysRevC.83.044618}.
\bibitem[{Niwase et~al.(2023)Niwase, Watanabe, Hirayama, Mukai, Schury,
  Andreyev, Hashimoto, Iimura, Ishiyama, Ito, Jeong, Kaji, Kimura, Miyatake,
  Morimoto, Moon, Oyaizu, Rosenbusch, Taniguchi, and Wada}]{NiwasePRL2023}
\bibinfo{author}{T.~Niwase}, \bibinfo{author}{Y.~X. Watanabe},
  \bibinfo{author}{Y.~Hirayama}, \bibinfo{author}{M.~Mukai},
  \bibinfo{author}{P.~Schury}, \bibinfo{author}{A.~N. Andreyev},
  \bibinfo{author}{T.~Hashimoto}, \bibinfo{author}{S.~Iimura},
  \bibinfo{author}{H.~Ishiyama}, \bibinfo{author}{Y.~Ito},
  \bibinfo{author}{S.~C. Jeong}, \bibinfo{author}{D.~Kaji},
  \bibinfo{author}{S.~Kimura}, \bibinfo{author}{H.~Miyatake},
  \bibinfo{author}{K.~Morimoto}, \bibinfo{author}{J.-Y. Moon},
  \bibinfo{author}{M.~Oyaizu}, \bibinfo{author}{M.~Rosenbusch},
  \bibinfo{author}{A.~Taniguchi}, \bibinfo{author}{M.~Wada},
  \bibinfo{journal}{Phys. Rev. Lett.} \bibinfo{volume}{130}
  (\bibinfo{year}{2023}) \bibinfo{pages}{132502}.
  \DOIprefix\doi{10.1103/PhysRevLett.130.132502}.
\bibitem[{Huang et~al.(2017)Huang, Tian, Wang, Wang, Gan, Liu, Yang, Ma, and
  Zhou}]{HuangNPR2017}
\bibinfo{author}{W.~X. Huang}, \bibinfo{author}{Y.~L. Tian},
  \bibinfo{author}{Y.~S. Wang}, \bibinfo{author}{J.~Y. Wang},
  \bibinfo{author}{Z.~G. Gan}, \bibinfo{author}{Z.~Liu},
  \bibinfo{author}{J.~Yang}, \bibinfo{author}{X.~W. Ma}, \bibinfo{author}{X.~H.
  Zhou}, \bibinfo{journal}{Nucl. Phys. Rev.} \bibinfo{volume}{34}
  (\bibinfo{year}{2017}) \bibinfo{pages}{409}.
  \DOIprefix\doi{10.11804/NuclPhysRev.34.03.409}.
\bibitem[{Simenel(2010)}]{SimenelPRL2010}
\bibinfo{author}{C.~Simenel}, \bibinfo{journal}{Phys. Rev. Lett.}
  \bibinfo{volume}{105} (\bibinfo{year}{2010}) \bibinfo{pages}{192701}.
  \DOIprefix\doi{10.1103/PhysRevLett.105.192701}.
\bibitem[{Sekizawa and Yabana(2013)}]{SekizawaPRC2013}
\bibinfo{author}{K.~Sekizawa}, \bibinfo{author}{K.~Yabana},
  \bibinfo{journal}{Phys. Rev. C} \bibinfo{volume}{88} (\bibinfo{year}{2013})
  \bibinfo{pages}{014614}. \DOIprefix\doi{10.1103/PhysRevC.88.014614}.
\bibitem[{Zhao et~al.(2016)Zhao, Li, Zhang, Wang, Li, Shen, Wang, and
  Wu}]{ZhaokaiPRC2016}
\bibinfo{author}{K.~Zhao}, \bibinfo{author}{Z.~Li}, \bibinfo{author}{Y.~Zhang},
  \bibinfo{author}{N.~Wang}, \bibinfo{author}{Q.~Li},
  \bibinfo{author}{C.~Shen}, \bibinfo{author}{Y.~Wang},
  \bibinfo{author}{X.~Wu}, \bibinfo{journal}{Phys. Rev. C} \bibinfo{volume}{94}
  (\bibinfo{year}{2016}) \bibinfo{pages}{024601}.
  \DOIprefix\doi{10.1103/PhysRevC.94.024601}.
\bibitem[{Wen et~al.(2019)Wen, Lin, Li, Zhu, Zhang, Zhang, Jia, Yang, Ma, Sun,
  Wang, Zhong, Sun, Yang, and Xu}]{WenPRC2019}
\bibinfo{author}{P.~W. Wen}, \bibinfo{author}{C.~J. Lin},
  \bibinfo{author}{C.~Li}, \bibinfo{author}{L.~Zhu},
  \bibinfo{author}{F.~Zhang}, \bibinfo{author}{F.~S. Zhang},
  \bibinfo{author}{H.~M. Jia}, \bibinfo{author}{F.~Yang},
  \bibinfo{author}{N.~R. Ma}, \bibinfo{author}{L.~J. Sun},
  \bibinfo{author}{D.~X. Wang}, \bibinfo{author}{F.~P. Zhong},
  \bibinfo{author}{H.~H. Sun}, \bibinfo{author}{L.~Yang},
  \bibinfo{author}{X.~X. Xu}, \bibinfo{journal}{Phys. Rev. C}
  \bibinfo{volume}{99} (\bibinfo{year}{2019}) \bibinfo{pages}{034606}.
  \DOIprefix\doi{10.1103/PhysRevC.99.034606}.
\bibitem[{Jiang and Wang(2020)}]{JiangPRC2020}
\bibinfo{author}{X.~Jiang}, \bibinfo{author}{N.~Wang}, \bibinfo{journal}{Phys.
  Rev. C} \bibinfo{volume}{101} (\bibinfo{year}{2020}) \bibinfo{pages}{014604}.
  \DOIprefix\doi{10.1103/PhysRevC.101.014604}.
\bibitem[{Wu et~al.(2022)Wu, Guo, Liu, and Peng}]{WuPLB2022}
\bibinfo{author}{Z.~Wu}, \bibinfo{author}{L.~Guo}, \bibinfo{author}{Z.~Liu},
  \bibinfo{author}{G.~Peng}, \bibinfo{journal}{Phys. Lett. B}
  \bibinfo{volume}{825} (\bibinfo{year}{2022}) \bibinfo{pages}{136886}.
  \DOIprefix\doi{https://doi.org/10.1016/j.physletb.2022.136886}.
\bibitem[{Chen et~al.(2022)Chen, Geng, Zeng, and Feng}]{ChenpenghuiPRC2022}
\bibinfo{author}{P.-H. Chen}, \bibinfo{author}{C.~Geng}, \bibinfo{author}{X.-H.
  Zeng}, \bibinfo{author}{Z.-Q. Feng}, \bibinfo{journal}{Phys. Rev. C}
  \bibinfo{volume}{106} (\bibinfo{year}{2022}) \bibinfo{pages}{054601}.
  \DOIprefix\doi{10.1103/PhysRevC.106.054601}.
\bibitem[{Feng(2023)}]{FengzhaoqingPRC2023}
\bibinfo{author}{Z.-Q. Feng}, \bibinfo{journal}{Phys. Rev. C}
  \bibinfo{volume}{108} (\bibinfo{year}{2023}) \bibinfo{pages}{L051601}.
  \DOIprefix\doi{10.1103/PhysRevC.108.L051601}.
\bibitem[{Liao et~al.(2024)Liao, Gao, Yang, Zhu, and Su}]{LiaoPRC2024}
\bibinfo{author}{Z.~Liao}, \bibinfo{author}{Z.~Gao}, \bibinfo{author}{Y.~Yang},
  \bibinfo{author}{L.~Zhu}, \bibinfo{author}{J.~Su}, \bibinfo{journal}{Phys.
  Rev. C} \bibinfo{volume}{109} (\bibinfo{year}{2024}) \bibinfo{pages}{054612}.
  \DOIprefix\doi{10.1103/PhysRevC.109.054612}.
\bibitem[{Zhao et~al.(2024)Zhao, Zhu, Zhang, and Bao}]{ZhaotianliangPRC2024}
\bibinfo{author}{T.~L. Zhao}, \bibinfo{author}{S.~H. Zhu},
  \bibinfo{author}{H.~F. Zhang}, \bibinfo{author}{X.~J. Bao},
  \bibinfo{journal}{Phys. Rev. C} \bibinfo{volume}{110} (\bibinfo{year}{2024})
  \bibinfo{pages}{064609}. \DOIprefix\doi{10.1103/PhysRevC.110.064609}.
\bibitem[{Zhang et~al.(2024{\natexlab{a}})Zhang, Vretenar, Nik\ifmmode
  \check{s}\else \v{s}\fi{}i\ifmmode~\acute{c}\else \'{c}\fi{}, Zhao, and
  Meng}]{ZhangPRC2024}
\bibinfo{author}{D.~D. Zhang}, \bibinfo{author}{D.~Vretenar},
  \bibinfo{author}{T.~Nik\ifmmode \check{s}\else
  \v{s}\fi{}i\ifmmode~\acute{c}\else \'{c}\fi{}}, \bibinfo{author}{P.~W. Zhao},
  \bibinfo{author}{J.~Meng}, \bibinfo{journal}{Phys. Rev. C}
  \bibinfo{volume}{109} (\bibinfo{year}{2024}{\natexlab{a}})
  \bibinfo{pages}{024614}. \DOIprefix\doi{10.1103/PhysRevC.109.024614}.
\bibitem[{Zhang et~al.(2024{\natexlab{b}})Zhang, Li, Vretenar, Nik{\v s}i{\'c},
  Ren, Zhao, and Meng}]{ZDDPRC2024}
\bibinfo{author}{D.~D. Zhang}, \bibinfo{author}{B.~Li},
  \bibinfo{author}{D.~Vretenar}, \bibinfo{author}{T.~Nik{\v s}i{\'c}},
  \bibinfo{author}{Z.~X. Ren}, \bibinfo{author}{P.~W. Zhao},
  \bibinfo{author}{J.~Meng}, \bibinfo{journal}{Phys. Rev. C}
  \bibinfo{volume}{109} (\bibinfo{year}{2024}{\natexlab{b}})
  \bibinfo{pages}{024316}. \DOIprefix\doi{10.1103/PhysRevC.109.024316}.
\bibitem[{Li et~al.(2024)Li, Li, Zhang, Tian, Wang, Guo, and
  Zhang}]{ChengLiPRC2024}
\bibinfo{author}{C.~Li}, \bibinfo{author}{T.~Li}, \bibinfo{author}{X.-R.
  Zhang}, \bibinfo{author}{J.~Tian}, \bibinfo{author}{N.~Wang},
  \bibinfo{author}{C.~Guo}, \bibinfo{author}{F.-S. Zhang},
  \bibinfo{journal}{Phys. Rev. C} \bibinfo{volume}{110} (\bibinfo{year}{2024})
  \bibinfo{pages}{044615}. \DOIprefix\doi{10.1103/PhysRevC.110.044615}.
\bibitem[{Zhang and Zhang(2025)}]{ZhangfengshouPRC2025}
\bibinfo{author}{G.~Zhang}, \bibinfo{author}{F.-S. Zhang},
  \bibinfo{journal}{Phys. Rev. C} \bibinfo{volume}{111} (\bibinfo{year}{2025})
  \bibinfo{pages}{014603}. \DOIprefix\doi{10.1103/PhysRevC.111.014603}.
\bibitem[{Feng et~al.(2025)Feng, Zhao, Liu, Wang, Zhang, and
  Zhang}]{FengPingPRC2025}
\bibinfo{author}{P.~Feng}, \bibinfo{author}{K.~Zhao}, \bibinfo{author}{Z.~Liu},
  \bibinfo{author}{N.~Wang}, \bibinfo{author}{F.-S. Zhang},
  \bibinfo{author}{Y.~Zhang}, \bibinfo{journal}{Phys. Rev. C}
  \bibinfo{volume}{111} (\bibinfo{year}{2025}) \bibinfo{pages}{024603}.
  \DOIprefix\doi{10.1103/PhysRevC.111.024603}.
\bibitem[{Ocal et~al.(2025)Ocal, Yilmaz, Arik, Ayik, and Umar}]{OcalPRC2025}
\bibinfo{author}{S.~E. Ocal}, \bibinfo{author}{O.~Yilmaz},
  \bibinfo{author}{M.~Arik}, \bibinfo{author}{S.~Ayik}, \bibinfo{author}{A.~S.
  Umar}, \bibinfo{journal}{Phys. Rev. C} \bibinfo{volume}{111}
  (\bibinfo{year}{2025}) \bibinfo{pages}{054613}.
  \DOIprefix\doi{10.1103/PhysRevC.111.054613}.
\bibitem[{Gao et~al.(2025)Gao, Sekizawa, and Zhu}]{GaoPRC2025}
\bibinfo{author}{Z.~Gao}, \bibinfo{author}{K.~Sekizawa},
  \bibinfo{author}{L.~Zhu}, \bibinfo{journal}{Phys. Rev. C}
  \bibinfo{volume}{112} (\bibinfo{year}{2025}) \bibinfo{pages}{014602}.
  \DOIprefix\doi{10.1103/zz3y-22fh}.
\bibitem[{Liao et~al.(2023)Liao, Zhu, Gao, Su, and Li}]{LiaoPRR2023}
\bibinfo{author}{Z.~Liao}, \bibinfo{author}{L.~Zhu}, \bibinfo{author}{Z.~Gao},
  \bibinfo{author}{J.~Su}, \bibinfo{author}{C.~Li}, \bibinfo{journal}{Phys.
  Rev. Res.} \bibinfo{volume}{5} (\bibinfo{year}{2023})
  \bibinfo{pages}{L022021}. \DOIprefix\doi{10.1103/PhysRevResearch.5.L022021}.
\bibitem[{Scamps(2024)}]{ScampsPRC2024}
\bibinfo{author}{G.~Scamps}, \bibinfo{journal}{Phys. Rev. C}
  \bibinfo{volume}{110} (\bibinfo{year}{2024}) \bibinfo{pages}{054605}.
  \DOIprefix\doi{10.1103/PhysRevC.110.054605}.
\bibitem[{Liao et~al.(2025)Liao, Gao, Yang, Fang, Zhu, and Su}]{LiaoPRC2025}
\bibinfo{author}{Z.~Liao}, \bibinfo{author}{Z.~Gao}, \bibinfo{author}{Y.~Yang},
  \bibinfo{author}{Y.~Fang}, \bibinfo{author}{L.~Zhu}, \bibinfo{author}{J.~Su},
  \bibinfo{journal}{Phys. Rev. C} \bibinfo{volume}{111} (\bibinfo{year}{2025})
  \bibinfo{pages}{024605}. \DOIprefix\doi{10.1103/PhysRevC.111.024605}.
\bibitem[{Le et~al.(2025)Le, Xing, Guo, and Wang}]{LePRC2025}
\bibinfo{author}{X.-K. Le}, \bibinfo{author}{F.-Z. Xing},
  \bibinfo{author}{S.-Q. Guo}, \bibinfo{author}{N.~Wang},
  \bibinfo{journal}{Phys. Rev. C} \bibinfo{volume}{111} (\bibinfo{year}{2025})
  \bibinfo{pages}{024618}. \DOIprefix\doi{10.1103/PhysRevC.111.024618}.
\bibitem[{et. al.(2021)}]{WilsonNature2021}
\bibinfo{author}{J.~N.~W. et. al.}, \bibinfo{journal}{Nature}
  \bibinfo{volume}{590} (\bibinfo{year}{2021}) \bibinfo{pages}{566--570}.
  \DOIprefix\doi{10.1038/s41586-021-03304-w}.
\bibitem[{Randrup and Vogt(2021)}]{RandrupPRL2021}
\bibinfo{author}{J.~Randrup}, \bibinfo{author}{R.~Vogt},
  \bibinfo{journal}{Phys. Rev. Lett.} \bibinfo{volume}{127}
  (\bibinfo{year}{2021}) \bibinfo{pages}{062502}.
  \DOIprefix\doi{10.1103/PhysRevLett.127.062502}.
\bibitem[{Vogt and Randrup(2021)}]{VogtPRC2021}
\bibinfo{author}{R.~Vogt}, \bibinfo{author}{J.~Randrup},
  \bibinfo{journal}{Phys. Rev. C} \bibinfo{volume}{103} (\bibinfo{year}{2021})
  \bibinfo{pages}{014610}. \DOIprefix\doi{10.1103/PhysRevC.103.014610}.
\bibitem[{Bulgac et~al.(2021)Bulgac, Abdurrahman, Jin, Godbey, Schunck, and
  Stetcu}]{BulgacPRL2021}
\bibinfo{author}{A.~Bulgac}, \bibinfo{author}{I.~Abdurrahman},
  \bibinfo{author}{S.~Jin}, \bibinfo{author}{K.~Godbey},
  \bibinfo{author}{N.~Schunck}, \bibinfo{author}{I.~Stetcu},
  \bibinfo{journal}{Phys. Rev. Lett.} \bibinfo{volume}{126}
  (\bibinfo{year}{2021}) \bibinfo{pages}{142502}.
  \DOIprefix\doi{10.1103/PhysRevLett.126.142502}.
\bibitem[{Bulgac et~al.(2022)Bulgac, Abdurrahman, Godbey, and
  Stetcu}]{BulgacPRL2022}
\bibinfo{author}{A.~Bulgac}, \bibinfo{author}{I.~Abdurrahman},
  \bibinfo{author}{K.~Godbey}, \bibinfo{author}{I.~Stetcu},
  \bibinfo{journal}{Phys. Rev. Lett.} \bibinfo{volume}{128}
  (\bibinfo{year}{2022}) \bibinfo{pages}{022501}.
  \DOIprefix\doi{10.1103/PhysRevLett.128.022501}.
\bibitem[{Scamps et~al.(2023)Scamps, Abdurrahman, Kafker, Bulgac, and
  Stetcu}]{ScampsPRC2023}
\bibinfo{author}{G.~Scamps}, \bibinfo{author}{I.~Abdurrahman},
  \bibinfo{author}{M.~Kafker}, \bibinfo{author}{A.~Bulgac},
  \bibinfo{author}{I.~Stetcu}, \bibinfo{journal}{Phys. Rev. C}
  \bibinfo{volume}{108} (\bibinfo{year}{2023}) \bibinfo{pages}{L061602}.
  \DOIprefix\doi{10.1103/PhysRevC.108.L061602}.
\bibitem[{Ren et~al.(2020{\natexlab{a}})Ren, Zhao, and Meng}]{RenPRC2020}
\bibinfo{author}{Z.~X. Ren}, \bibinfo{author}{P.~W. Zhao},
  \bibinfo{author}{J.~Meng}, \bibinfo{journal}{Phys. Rev. C}
  \bibinfo{volume}{102} (\bibinfo{year}{2020}{\natexlab{a}})
  \bibinfo{pages}{044603}. \DOIprefix\doi{10.1103/PhysRevC.102.044603}.
\bibitem[{Ren et~al.(2020{\natexlab{b}})Ren, Zhao, and Meng}]{RenPLB2020}
\bibinfo{author}{Z.~X. Ren}, \bibinfo{author}{P.~W. Zhao},
  \bibinfo{author}{J.~Meng}, \bibinfo{journal}{Phys. Lett. B}
  \bibinfo{volume}{801} (\bibinfo{year}{2020}{\natexlab{b}})
  \bibinfo{pages}{135194}. \DOIprefix\doi{10.1016/j.physletb.2019.135194}.
\bibitem[{Ren et~al.(2022{\natexlab{a}})Ren, Zhao, Vretenar, Nik{\v s}i{\'c},
  Zhao, and Meng}]{RenPRC2022}
\bibinfo{author}{Z.~X. Ren}, \bibinfo{author}{J.~Zhao},
  \bibinfo{author}{D.~Vretenar}, \bibinfo{author}{T.~Nik{\v s}i{\'c}},
  \bibinfo{author}{P.~W. Zhao}, \bibinfo{author}{J.~Meng},
  \bibinfo{journal}{Phys. Rev. C} \bibinfo{volume}{105}
  (\bibinfo{year}{2022}{\natexlab{a}}) \bibinfo{pages}{044313}.
  \DOIprefix\doi{10.1103/PhysRevC.105.044313}.
\bibitem[{Ren et~al.(2022{\natexlab{b}})Ren, Zhao, and
  Meng}]{RenPRC2022-chiral}
\bibinfo{author}{Z.~X. Ren}, \bibinfo{author}{P.~W. Zhao},
  \bibinfo{author}{J.~Meng}, \bibinfo{journal}{Phys. Rev. C}
  \bibinfo{volume}{105} (\bibinfo{year}{2022}{\natexlab{b}})
  \bibinfo{pages}{L011301}. \DOIprefix\doi{10.1103/PhysRevC.105.L011301}.
\bibitem[{Ren et~al.(2022{\natexlab{c}})Ren, Vretenar, Nik{\v s}i{\'c}, Zhao,
  Zhao, and Meng}]{RenPRL2022}
\bibinfo{author}{Z.~X. Ren}, \bibinfo{author}{D.~Vretenar},
  \bibinfo{author}{T.~Nik{\v s}i{\'c}}, \bibinfo{author}{P.~W. Zhao},
  \bibinfo{author}{J.~Zhao}, \bibinfo{author}{J.~Meng}, \bibinfo{journal}{Phys.
  Rev. Lett.} \bibinfo{volume}{128} (\bibinfo{year}{2022}{\natexlab{c}})
  \bibinfo{pages}{172501}. \DOIprefix\doi{10.1103/PhysRevLett.128.172501}.
\bibitem[{Li et~al.(2023)Li, Vretenar, Ren, Nik{\v s}i{\'c}, Zhao, Zhao, and
  Meng}]{LiPRC2023}
\bibinfo{author}{B.~Li}, \bibinfo{author}{D.~Vretenar}, \bibinfo{author}{Z.~X.
  Ren}, \bibinfo{author}{T.~Nik{\v s}i{\'c}}, \bibinfo{author}{J.~Zhao},
  \bibinfo{author}{P.~W. Zhao}, \bibinfo{author}{J.~Meng},
  \bibinfo{journal}{Phys. Rev. C} \bibinfo{volume}{107} (\bibinfo{year}{2023})
  \bibinfo{pages}{014303}. \DOIprefix\doi{10.1103/PhysRevC.107.014303}.
\bibitem[{Li et~al.(2024{\natexlab{a}})Li, Vretenar, Nik{\v s}i{\'c}, Zhao, and
  Meng}]{LiPRC2024-fission}
\bibinfo{author}{B.~Li}, \bibinfo{author}{D.~Vretenar},
  \bibinfo{author}{T.~Nik{\v s}i{\'c}}, \bibinfo{author}{P.~W. Zhao},
  \bibinfo{author}{J.~Meng}, \bibinfo{journal}{Phys. Rev. C}
  \bibinfo{volume}{110} (\bibinfo{year}{2024}{\natexlab{a}})
  \bibinfo{pages}{034302}. \DOIprefix\doi{10.1103/PhysRevC.110.034302}.
\bibitem[{Li et~al.(2024{\natexlab{b}})Li, Zhao, and Meng}]{LiPLB2024}
\bibinfo{author}{B.~Li}, \bibinfo{author}{P.~W. Zhao},
  \bibinfo{author}{J.~Meng}, \bibinfo{journal}{Phys. Lett. B}
  \bibinfo{volume}{856} (\bibinfo{year}{2024}{\natexlab{b}})
  \bibinfo{pages}{138877}. \DOIprefix\doi{10.1016/j.physletb.2024.138877}.
\bibitem[{Ma and Ma(2018)}]{MaPPNP2018}
\bibinfo{author}{C.-W. Ma}, \bibinfo{author}{Y.-G. Ma}, \bibinfo{journal}{Prog.
  Part. Nucl. Phys.} \bibinfo{volume}{99} (\bibinfo{year}{2018})
  \bibinfo{pages}{120--158}. \DOIprefix\doi{10.1016/j.ppnp.2018.01.002}.
\bibitem[{Zhao et~al.(2010)Zhao, Li, Yao, and Meng}]{ZhaoPRC2010}
\bibinfo{author}{P.~W. Zhao}, \bibinfo{author}{Z.~P. Li},
  \bibinfo{author}{J.~M. Yao}, \bibinfo{author}{J.~Meng},
  \bibinfo{journal}{Phys. Rev. C} \bibinfo{volume}{82} (\bibinfo{year}{2010})
  \bibinfo{pages}{054319}. \DOIprefix\doi{10.1103/PhysRevC.82.054319}.
\bibitem[{Ren et~al.(2017)Ren, Zhang, and Meng}]{RenPRC2017}
\bibinfo{author}{Z.~X. Ren}, \bibinfo{author}{S.~Q. Zhang},
  \bibinfo{author}{J.~Meng}, \bibinfo{journal}{Phys. Rev. C}
  \bibinfo{volume}{95} (\bibinfo{year}{2017}) \bibinfo{pages}{024313}.
  \DOIprefix\doi{10.1103/PhysRevC.95.024313}.
\bibitem[{Li et~al.(2020)Li, Ren, and Zhao}]{LiPRC2020}
\bibinfo{author}{B.~Li}, \bibinfo{author}{Z.~X. Ren}, \bibinfo{author}{P.~W.
  Zhao}, \bibinfo{journal}{Phys. Rev. C} \bibinfo{volume}{102}
  (\bibinfo{year}{2020}) \bibinfo{pages}{044307}.
  \DOIprefix\doi{10.1103/PhysRevC.102.044307}.
\bibitem[{Zhang et~al.(2022)Zhang, Ren, Zhao, Vretenar, Nik\ifmmode
  \check{s}\else \v{s}\fi{}i\ifmmode~\acute{c}\else \'{c}\fi{}, and
  Meng}]{ZDDPRC2022}
\bibinfo{author}{D.~D. Zhang}, \bibinfo{author}{Z.~X. Ren},
  \bibinfo{author}{P.~W. Zhao}, \bibinfo{author}{D.~Vretenar},
  \bibinfo{author}{T.~Nik\ifmmode \check{s}\else
  \v{s}\fi{}i\ifmmode~\acute{c}\else \'{c}\fi{}}, \bibinfo{author}{J.~Meng},
  \bibinfo{journal}{Phys. Rev. C} \bibinfo{volume}{105} (\bibinfo{year}{2022})
  \bibinfo{pages}{024322}. \DOIprefix\doi{10.1103/PhysRevC.105.024322}.
\bibitem[{Zhang(2023)}]{ZDDIJMPE2023}
\bibinfo{author}{D.~D. Zhang}, \bibinfo{journal}{Int. J. Mod. Phys. E}
  \bibinfo{volume}{32} (\bibinfo{year}{2023}) \bibinfo{pages}{2340009}.
  \DOIprefix\doi{10.1142/S0218301323400098}.
\bibitem[{Xu et~al.(2024{\natexlab{a}})Xu, Li, Ren, and Zhao}]{XuPRC2024}
\bibinfo{author}{F.~F. Xu}, \bibinfo{author}{B.~Li}, \bibinfo{author}{Z.~X.
  Ren}, \bibinfo{author}{P.~W. Zhao}, \bibinfo{journal}{Phys. Rev. C}
  \bibinfo{volume}{109} (\bibinfo{year}{2024}{\natexlab{a}})
  \bibinfo{pages}{014311}. \DOIprefix\doi{10.1103/PhysRevC.109.014311}.
\bibitem[{Xu et~al.(2024{\natexlab{b}})Xu, Wang, Wang, Ring, and
  Zhao}]{XuPRL2024}
\bibinfo{author}{F.~F. Xu}, \bibinfo{author}{Y.~K. Wang},
  \bibinfo{author}{Y.~P. Wang}, \bibinfo{author}{P.~Ring},
  \bibinfo{author}{P.~W. Zhao}, \bibinfo{journal}{Phys. Rev. Lett.}
  \bibinfo{volume}{133} (\bibinfo{year}{2024}{\natexlab{b}})
  \bibinfo{pages}{022501}. \DOIprefix\doi{10.1103/PhysRevLett.133.022501}.
\bibitem[{Xu et~al.(2024{\natexlab{c}})Xu, Li, Ring, and Zhao}]{XuPLB2024}
\bibinfo{author}{F.~F. Xu}, \bibinfo{author}{B.~Li}, \bibinfo{author}{P.~Ring},
  \bibinfo{author}{P.~W. Zhao}, \bibinfo{journal}{Phys. Lett. B}
  \bibinfo{volume}{856} (\bibinfo{year}{2024}{\natexlab{c}})
  \bibinfo{pages}{138893}. \DOIprefix\doi{10.1016/j.physletb.2024.138893}.
\bibitem[{Li et~al.(2024)Li, Vretenar, Nik\ifmmode \check{s}\else
  \v{s}\fi{}i\ifmmode~\acute{c}\else \'{c}\fi{}, Zhang, Zhao, and
  Meng}]{LiPRC2024}
\bibinfo{author}{B.~Li}, \bibinfo{author}{D.~Vretenar},
  \bibinfo{author}{T.~Nik\ifmmode \check{s}\else
  \v{s}\fi{}i\ifmmode~\acute{c}\else \'{c}\fi{}}, \bibinfo{author}{D.~D.
  Zhang}, \bibinfo{author}{P.~W. Zhao}, \bibinfo{author}{J.~Meng},
  \bibinfo{journal}{Phys. Rev. C} \bibinfo{volume}{110} (\bibinfo{year}{2024})
  \bibinfo{pages}{034611}. \DOIprefix\doi{10.1103/PhysRevC.110.034611}.
\bibitem[{Bass(1980)}]{Bass1980}
\bibinfo{author}{R.~Bass}, \bibinfo{title}{Nuclear Reactions with Heavy Ions},
  \bibinfo{publisher}{Springer-Verlag, Berlin}, \bibinfo{year}{1980}.
\bibitem[{Moretto and Schmitt(1980)}]{MorettoPRC1980}
\bibinfo{author}{L.~G. Moretto}, \bibinfo{author}{R.~P. Schmitt},
  \bibinfo{journal}{Phys. Rev. C} \bibinfo{volume}{21} (\bibinfo{year}{1980})
  \bibinfo{pages}{204--216}. \DOIprefix\doi{10.1103/PhysRevC.21.204}.
\bibitem[{Moretto et~al.(1989)Moretto, Peaslee, and Wozniak}]{MorettoNPA1989}
\bibinfo{author}{L.~G. Moretto}, \bibinfo{author}{G.~F. Peaslee},
  \bibinfo{author}{G.~J. Wozniak}, \bibinfo{journal}{Nucl. Phys. A}
  \bibinfo{volume}{502} (\bibinfo{year}{1989}) \bibinfo{pages}{453--472}.
  \DOIprefix\doi{10.1016/0375-9474(89)90682-9}.

\end{thebibliography}

\end{document}